\documentclass[prl, twocolumn, floatfix, superscriptaddress, longbibliography, preprintnumbers]{revtex4-1}

\usepackage{amsmath, amssymb, bbm, bbold}
\usepackage{graphicx}
\usepackage{color}

\newcommand{\ket}[1]{{| #1 \rangle}}
\newcommand{\bra}[1]{{\langle #1 |}}
\newcommand{\ee}{{\rm e}}
\newcommand{\ii}{{\it i}}
\newcommand{\dd}{{\rm d}}
\newcommand{\Tr}{{\rm Tr}}
\newcommand{\sinc}{{\rm sinc}}

\begin{document}

\title{Disorder-induced dephasing in backscattering-free quantum transport}

\author{Clemens Gneiting}
\email{clemens.gneiting@riken.jp}
\affiliation{Quantum Condensed Matter Research Group, RIKEN, Wako-shi, Saitama 351-0198, Japan}
\author{Franco Nori}
\email{fnori@riken.jp}
\affiliation{Quantum Condensed Matter Research Group, RIKEN, Wako-shi, Saitama 351-0198, Japan}
\affiliation{Department of Physics, University of Michigan, Ann Arbor, Michigan 48109-1040, USA}

\date{\today}

\begin{abstract}
We analyze the disorder-perturbed transport of quantum states in the absence of backscattering. This comprises, for instance, the propagation of edge-mode wave packets in topological insulators, or the propagation of photons in inhomogeneous media. We quantify the disorder-induced dephasing, which we show to be bound. Moreover, we identify a gap condition to remain in the backscattering-free regime despite disorder-induced momentum broadening. Our analysis comprises the full disorder-averaged quantum state, on the level of both populations and coherences, appreciating states as potential carriers of quantum information. The well-definedness of states is guaranteed by our treatment of the nonequilibrium dynamics with Lindblad master equations.
\end{abstract}

\preprint{\textsf{published in Phys.~Rev.~Lett.~{\bf 119}, 176802 (2017)}}

\maketitle

\section{Introduction}

Among the most distinct characteristics of topological insulators are the existence of chiral edge modes and their robust transport behavior, reflected by the absence of backscattering even in the presence of disorder. Their remarkable features make them potential candidates for technological innovations such as, for example, electronic devices with low power consumption, or, in combination with an inherent spin-current correlation, spintronics devices (for reviews on topological insulators, see Refs.~\cite{Hasan2010colloquium, Qi2011topological} and references therein).

\begin{figure}[htb]
	\includegraphics[width=0.99\columnwidth]{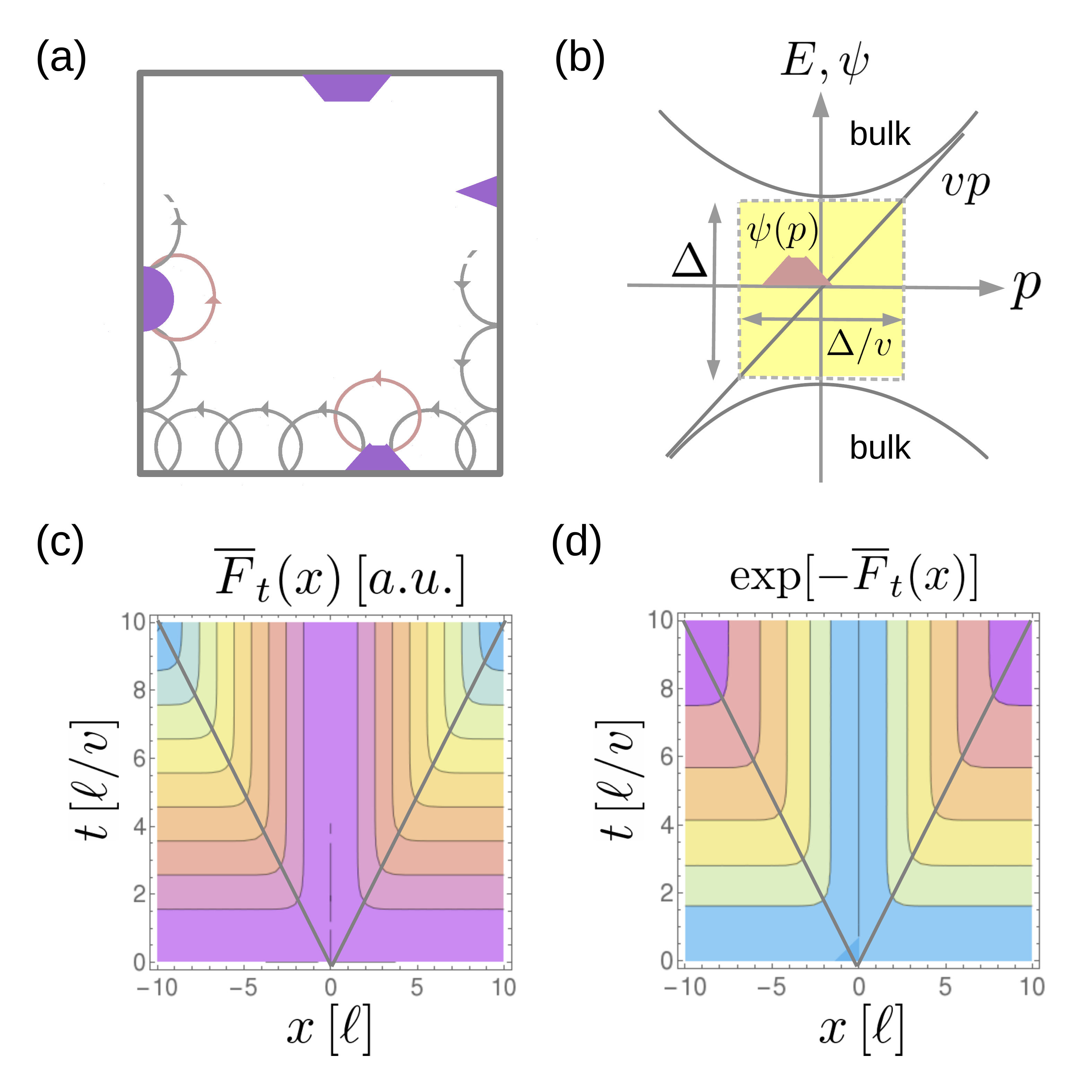}
	\caption{\label{Fig:backscattering_free_transport} Disorder-induced dephasing in backscattering-free propagation. (a) Quasiclassical representation of skipping orbits in a quantum Hall sample in the presence of a strong magnetic field. Because of the directional Lorentz-force bending of the orbits, obstacles/impurities cannot reverse the direction of motion. (b) Generic band model of a topological insulator. Edge modes exist in the bulk band gap of width $\Delta$. To propagate backscattering-free, edge mode wave packets must remain confined in the gap region. (c) and (d) Decoherence cone describing the spatiotemporal dephasing behavior of the disorder-averaged quantum state [cf.~Eq.~(\ref{Eq:master_equation_solution})]. Coherences within the cone remain unaffected, outside they decay. Values in (c) increase from $0$ (violet).}
\end{figure}

Their robust transport properties render topological insulators also attractive for more advanced applications, e.g., in quantum information processing. In analogy to photonic systems, possible applications could, for instance, encode quantum information in sequences of subsequently propagating edge states. Such schemes have been successfully employed for quantum communication tasks \cite{Brendel1999pulsed, Thew2002experimental}, and can allow for universal quantum computation with basic linear optics elements \cite{Knill2001scheme}.

While edge states are insensitive to disorder in their transport properties, they are, however, not immune to disorder effects when it comes to dephasing, reflected by a disorder-induced deformation of states. This poses a potential obstacle to their successful deployment as carriers of quantum information, where processing units, such as beam splitters, ideally require modes that perfectly coincide when matched. The harnessing of topological insulators (and alike systems) for such purposes thus requires a careful analysis of {\it disorder-induced dephasing} and its consequences. The same holds for photons, which fluctuate when propagating in an inhomogeneous medium.

In this article, we conduct such analysis. To this end, we solve the disorder-perturbed time evolution of backscattering-freely propagating quantum states. This is achieved by employing the recently established treatment of disordered quantum systems with Lindblad master equations \cite{Gneiting2016incoherent, Kropf2016effective, Gneiting2017quantum, Chen2017simulating}. It covers the disorder impact on the level of both, the populations and coherences of the disorder-averaged state, as required for a statistically robust dephasing analysis; at the same time, it certifies that the quantum evolution is at all times welldefined and physical \cite{Lindblad1976generators, Breuer2002theory}. We quantify the disorder-induced, spatiotemporal dephasing behavior, which results in a bounded decay of the purity of the ensemble-averaged state, and in a finite broadening of its momentum distribution. While the former is relevant to assess the impact of disorder on processing units, such as, e.g., beam splitters, the latter allows us to formulate a condition on the disorder for edge states, on the basis that the latter should remain confined in the backscattering-free bulk band gap.

\section{Effective evolution}

We consider the backscattering-free propagation of a single, spinless quantum particle in one dimension. This describes, for example, the propagation of edge mode wave packets along the edge in a 2D sample (Fig.~\ref{Fig:backscattering_free_transport}). In order to keep our analysis generic, we do not model internal or transversal degrees of freedom, or the mechanism for the robustness. In the case of chiral edge modes (which typically decay exponentially into the bulk), the latter is of topological origin, resulting in the nonexistence of counterpropagating modes \cite{Hasan2010colloquium, Qi2011topological}. The characterizing feature is a unidirectional drift, even in the presence of a (weak) disorder potential. If the disorder strength exceeds a tolerance threshold, however, backscattering sets in. In the case of chiral edge modes, this would be if the state is scattered into the bulk [Fig.~\ref{Fig:backscattering_free_transport}(b)]; in the case of photons, it occurs at the interface between two optically distinct media.

We assume that the drift velocity $v$ is, to leading order, constant. This is usually an excellent approximation, and in some generic model systems even exact \cite{Jackiw1976solitons, Qi2011topological, Hasan2010colloquium}. On the level of the time evolution, such constant drift is captured by translation operators, or, in terms of the Hamiltonian, by their generators, (quasi-)momentum operators, giving rise to the Hamiltonian $\hat{H}_{\varepsilon} = v \hat{p} + V_{\varepsilon}(\hat{x})$, with $x$ the direction along the edge and $v>0$. The (multi-)index $\varepsilon$ labels different disorder realizations, occurring with probability $p_{\varepsilon}$ (for simplicity we write integrals throughout, e.g., $\int \dd \varepsilon \, p_{\varepsilon} = 1$).

The disorder potential $V_{\varepsilon}(\hat{x}) = \int_{-\infty}^{\infty} \dd x \, V_{\varepsilon}(x) \ket{x}\bra{x}$ may be homogeneous on average, displaying translation-covariant two-point correlations, $\int \dd \varepsilon \, p_{\varepsilon} \, V_{\varepsilon}(x) V_{\varepsilon}(x') \equiv C(x-x') = \int_{-\infty}^{\infty} \dd q \, \ee^{\frac{\ii}{\hbar} q (x-x')} G(q)$, characterized by the momentum transfer distribution $G(q)$ (see also Refs.~\cite{Gneiting2016incoherent, Gneiting2017quantum}). For simplicity, the disorder potential may, in addition, vanish on average, $\int \dd \varepsilon \, p_{\varepsilon} \, V_{\varepsilon}(x) = 0$, i.e., the average Hamiltonian is given by $\hat{\overline{H}} \equiv \int \dd \varepsilon \, p_{\varepsilon} \, \hat{H}_{\varepsilon} = v \hat{p}$. We assume that the disorder potential is weak, in the sense that, due to the drift, positive and negative potential variations rapidly compensate. This will become clear below.

The dynamics of the disorder-averaged state $\overline{\rho}(t) = \int \dd \varepsilon \, p_{\varepsilon} \ee^{-\frac{\ii}{\hbar} \hat{H}_{\varepsilon} t} \rho_0 \ee^{\frac{\ii}{\hbar} \hat{H}_{\varepsilon} t}$ is not characterized by a Hamiltonian alone, but must, in general, be described by a quantum master equation \cite{Gneiting2016incoherent, Kropf2016effective, Gneiting2017quantum}. In Ref.~\cite{Gneiting2017quantum}, a Lindblad master equation for general disorder configurations, which is perturbative to second order in the disorder potential, is derived and applied to the disorder-perturbed propagation in parabolic bands. Here, we evaluate it for the backscattering-free propagation of chiral edge states, subject to a weak, homogeneous disorder potential. We obtain
\begin{subequations} \label{Eq:perturbative_master_equation_in_homogeneous_disorder}
	\begin{align}
		\partial_t \overline{\rho}(t) =& -\frac{\ii}{\hbar} [\hat{H}_{\rm eff}(t), \overline{\rho}(t)] \nonumber \\
		& +\sum_{\alpha \in \{ \pm 1 \}} \frac{2 \alpha}{\hbar^2} \int_{-\infty}^{\infty} \dd q \, G(q) \int_{0}^{t} \dd t' \mathcal{L}\big(\hat{L}_{q, t'}^{(\alpha)}, \overline{\rho}(t)\big) ,
	\end{align}
where $\mathcal{L}(\hat{L}, \rho) \equiv \hat{L} \rho \hat{L}^{\dagger} - \frac{1}{2} \hat{L}^{\dagger} \hat{L} \rho - \frac{1}{2} \rho \hat{L}^{\dagger} \hat{L}$. Note that we have already absorbed the disorder integral in the correlation function and exploited the translation invariance of the latter, giving rise to a reformulation in terms of the momentum transfer distribution $G(q)$ (see also Refs.~\cite{Gneiting2016incoherent, Gneiting2017quantum}). The effective Hamiltonian $H_{\rm eff}(t) = H_{\rm eff}^{\dagger}(t)$ and Lindblad operators $\hat{L}_{q, t}^{(\alpha)}$ are given by
	\begin{align}
		\hat{H}_{\rm eff}(t) &= \hat{\overline{H}} -\frac{\ii}{2 \hbar} \int_{-\infty}^{\infty} \dd q \, G(q) \int_{0}^{t} \dd t' \, [\hat{V}_{q}, \hat{\tilde{V}}_{-q}(t')] , \nonumber \\
		\hat{L}_{q, t}^{(\alpha)} &= \frac{1}{2} \big[ \hat{V}_{q} + \alpha \, \hat{\tilde{V}}_{q}(t) \big] ,
	\end{align}
\end{subequations}
where $\hat{V}_q = \ee^{\frac{\ii}{\hbar} q \hat{x}}$ (describing momentum kicks) and $\hat{\tilde{V}}_q(t) = \ee^{-\frac{\ii}{\hbar} \hat{\overline{H}} t} \hat{V}_q \ee^{\frac{\ii}{\hbar} \hat{\overline{H}} t}$. Note that the Lindblad operators $\hat{L}_{q, t}^{(\alpha)}$ are not Hermitian, in contrast to the Lindblad operators $\hat{L}_{\varepsilon, t}^{(\alpha)}$ of the general master equation \cite{Gneiting2017quantum}.

Since we have not yet specified $\hat{\overline{H}}$ in Eq.~(\ref{Eq:perturbative_master_equation_in_homogeneous_disorder}), it still describes propagation in homogeneous disorder for arbitrary kinetic terms. With $\hat{\overline{H}} = v \hat{p}$, one obtains $\hat{\tilde{V}}_q(t) = \ee^{\frac{\ii}{\hbar} q \hat{x}} \ee^{-\frac{\ii}{\hbar} v q t}$, i.e., $\hat{V}_q$ is merely modified by a time- and momentum-dependent phase factor. In this case, Eq.~(\ref{Eq:perturbative_master_equation_in_homogeneous_disorder}) can be significantly simplified, resulting in a manifestly translation-covariant master equation \cite{Kossakowski1972quantum, Manita1991properties, Botvich1991translation, Holevo1995translation}:
\begin{align} \label{Eq:master_equation_evaluated}
\partial_t& \overline{\rho}(t) = -\frac{\ii}{\hbar} [v \hat{p}, \overline{\rho}(t)] \\
 &+ \int_{-\infty}^{\infty} \!\!\! \dd q \; \frac{2 t G(q)}{\hbar^2} \sinc \! \left[ \frac{q v t}{\hbar} \right] \left\{ \ee^{\frac{\ii}{\hbar} q \hat{x}} \overline{\rho}(t) \ee^{-\frac{\ii}{\hbar} q \hat{x}} - \overline{\rho}(t) \right\} . \nonumber
\end{align}
A similar master equation is familiar from the context of collisional decoherence, there, however, with time-constant decoherence rates \cite{Gallis1990environmental, Hornberger2003collisional}. Here, we obtain temporally oscillating rates, which, as we show next, give rise to a substantially refined spatiotemporal, disorder-induced dephasing behavior. In particular, and in contrast to the short-time limit discussed in Ref.~\cite{Gneiting2016incoherent}, the disorder-induced dephasing remains bounded.

The master equation (\ref{Eq:master_equation_evaluated}) can be solved exactly and the solution reads, in the position representation [$G(-q) = G(q)$],
\begin{subequations} \label{Eq:master_equation_solution}
\begin{align}
\bra{x} \overline{\rho}(t) \ket{x'} = \bra{x-v t} \rho_0 \ket{x'-v t} \exp \left[ -\overline{F}_t(x-x') \right] ,
\end{align}
with the disorder influence summarized by
\begin{align} \label{Eq:disorder_influence}
\overline{F}_t(x) = \frac{t^2}{\hbar^2} \int \dd q \, G(q) \, \sinc^2 \left[ \frac{q v t}{2 \hbar} \right] \left\{ 1-\cos \left[ \frac{q x}{\hbar} \right] \right\} .
\end{align}
\end{subequations}
We emphasize that this solution holds for arbitrary initial states $\rho_0$ and arbitrary correlations $C(x)$. It describes the drift of the initial state with velocity $v$, along with a spatiotemporally resolved decay of the coherences, following the pattern of a decoherence cone [Figs.~\ref{Fig:backscattering_free_transport}(c) and \ref{Fig:backscattering_free_transport}(d)]. Colloquially speaking, the coherence loss between two points scales with their spatial separation $\Delta x$, terminating after $v t > \Delta x$. Concretely, for Gaussian correlations,
\begin{align} \label{Eq:Gaussian_correlations}
C(x) = C_0 \, \exp \left[- \left(\frac{x}{\ell} \right)^2 \right] ,
\end{align}
where $\ell$ denotes the correlation length, one obtains, with $G(q) \equiv \frac{1}{2 \pi \hbar} \int_{-\infty}^{\infty} \dd x \, \ee^{-\frac{\ii}{\hbar} q x} C(x) = \frac{C_0 \ell}{2 \sqrt{\pi} \hbar} \ee^{-\frac{1}{4}\left( \frac{q \ell}{\hbar} \right)^2}$, the disorder influence $\overline{F}_t(x) = \frac{C_0 \ell^2}{\hbar^2 v^2} \left\{ 2 \overline{f} \left( \frac{v t}{\ell} \right) + 2 \overline{f} \left( \frac{x}{\ell} \right) - \overline{f} \left( \frac{x - v t}{\ell} \right) - \overline{f} \left( \frac{x + v t}{\ell} \right) - 2 \overline{f}(0) \right\}$, with $\overline{f}(x) = x \, {\rm erf}(x) + (\ee^{-x^2}/\sqrt{\pi})$.

We focus here on Gaussian correlations, since they may be used generically to model many physical situations. Other correlation behavior is also conceivable, for instance of the Ornstein-Uhlenbeck type. In the limit of vanishing correlation length, as described by $\delta$ correlations, $C(x) = C_0 \delta(x)$, with $G(q) = \frac{C_0}{2 \pi \hbar}$, one obtains $\overline{F}_t^{(\delta)}(x) = \frac{C_0}{v^2 \hbar^2} [|v t| \, \Theta(|x|-|v t|) + |x| \, \Theta(|v t|-|x|)]$, with $\Theta(x)$ the unit-step function.

Often it is convenient to work in quantum phase space \cite{ Gneiting2017quantum, Gneiting2013quantum}. In terms of the characteristic function $\overline{\chi}_t(s,q) = \int \dd x \dd p \, \ee^{-\frac{\ii}{\hbar}(q x-p s)} \overline{W}_t(x,p)$ of the Wigner function $\overline{W}_t(x,p) = \frac{1}{2 \pi \hbar} \int \dd x' \ee^{\frac{\ii}{\hbar} p x'} \bra{x-\frac{x'}{2}} \overline{\rho}(t) \ket{x+\frac{x'}{2}}$, the solution then reads $\overline{\chi}_t(s,q) = \chi_0(s,q) \ee^{-\frac{\ii}{\hbar} q v t} \ee^{-\overline{F}_t(s)}$.

In the remainder, we often assume, to be generic, Gaussian initial states: $\psi_0(x) = \exp \left[-\frac{1}{4}(\frac{x}{\sigma})^2 + \frac{\ii}{\hbar} p_0 x \right]/\sqrt{\sqrt{2 \pi} \sigma}$, or, in phase space, $\chi_0(s,q) = \exp \left[ -\frac{1}{8}(\frac{s}{\sigma})^2 - \frac{1}{2} (\frac{q \, \sigma}{\hbar})^2 + \frac{\ii}{\hbar} p_0 s \right]$. Note that, here, $p_0$ lacks its usual kinetic interpretation, as the spatial displacement is completely determined by the drift velocity $v$. Nevertheless, it determines the position of the wave packet in the band [Fig.~\ref{Fig:backscattering_free_transport}(b)].

Accordingly, the position expectation values, when evaluated for (\ref{Eq:master_equation_solution}), read $\langle \hat{x} \rangle(t) = v t$, irrespective of the presence of a disorder potential, and in contrast to the propagation in a disordered parabolic dispersion band \cite{Gneiting2017quantum}. Similarly, the position variance remains timeinvariant, unaffected by both disorder and dispersion; in case of the above Gaussian initial state $\langle (\hat{x}-\langle \hat{x} \rangle)^2 \rangle(t) = \sigma^2$. However, as we will show below, a disorder-induced broadening of the momentum distribution persists in dispersion-free propagation.

\section{Disorder-induced dephasing}

The same initial state evolves differently in different disorder realizations, giving rise to disorder-induced dephasing. The latter is reflected by a loss of purity of the disorder-averaged state as compared to the initial state. This purity can thus be employed to assess the disorder-induced divergence among states, or, for that matter, to assess the deviation of disorder-perturbed states from the unperturbed evolution, this way quantifying the disorder impact.

If we evaluate the purity $r(t) \equiv \Tr[\rho(t)^2]$ for the solution~(\ref{Eq:master_equation_solution}), a Gaussian initial state, and Gaussian correlations~(\ref{Eq:Gaussian_correlations}), we obtain $r(t) = 1 - \frac{2 \ell^2 C_0}{v^2 \hbar^2} \Big\{ \sqrt{1 + 4 \left(\frac{\sigma}{\ell} \right)^2} \left[ 1 - \ee^{-(v t)^2/(\ell^2+4 \sigma^2)} \right] - \left[ 1 - \ee^{-(v t/\ell)^2} \right] + \sqrt{\pi} \frac{v t}{\ell} \left( {\rm erf}[\frac{v t}{\ell}] - {\rm erf}[\frac{v t}{\sqrt{\ell^2 + 4 \sigma^2}}] \right) \Big\}$, where we assumed small purity losses. In the limit $v t \gg \sqrt{\ell^2 + 4 \sigma^2}$, this reduces to
\begin{align} \label{Eq:purity_evolution}
r(t \gg \sqrt{\ell^2 + 4 \sigma^2}/v) = 1 - \frac{2 \ell^2 C_0}{v^2 \hbar^2} \left\{ \sqrt{1 + 4 \left(\frac{\sigma}{\ell} \right)^2} - 1 \right\} ,
\end{align}
i.e., the purity assumes a plateau value which is determined by both the characteristics of the disorder and the initial state. If $\sigma \ll \ell$ or $\sigma \gg \ell$, Eq.~(\ref{Eq:purity_evolution}) further simplifies to $r(t) = 1 - \frac{4 \sigma^2 C_0}{v^2 \hbar^2}$ or $r(t) = 1 - \frac{4 \sigma \ell C_0}{v^2 \hbar^2}$, respectively.

We find that, in contrast to the disorder-perturbed propagation in a parabolic band, which suffers an ongoing purity loss due to the dispersive spreading of the wave packet \cite{Gneiting2017quantum}, the purity plateau remains stable in the considered case of constant drift. In combination with the absence of backscattering, this (controllable) boundedness of the disorder-induced dephasing renders these systems promising as carriers of quantum information.

Note that, in the case of individual disorder realizations, the wave packet fluctuates as it propagates along the disorder potential. Comparing the wave packet at separations larger than the correlation length $\ell$, i.e., after the memory of the disorder potential is lost, is then equivalent to an ensemble average over different disorder realizations. In this sense, ensemble average and evolution in a single realization are connected by ergodicity.

For example, we now determine the impact of disorder on the functioning of beam splitters [Fig.~\ref{Fig:purity_evolution}(a)]. The latter are among the indispensable processing units in linear optics, quantum computation, quantum communication, and quantum foundations \cite{Knill2001scheme, Brendel1999pulsed, Aaronson2011computational, Hong1987measurement, Tichy2015double, Ra2017reversed}. Their optimal operation, based on constructive and destructive interference in the output arms (denoted as $\pm$), respectively, assumes identical input states, where a phase shift $\varphi$ in one input arm may determine the output probabilities, i.e., in the idealized case, ${\rm prob}_\pm(\varphi) = \frac{1}{2}(1 \pm \sin \varphi)$ (we assume balanced beam splitters). If, realistically, the input states $\ket{\psi}$ and $\ket{\psi'}$ slightly differ, e.g., due to different disorder histories, one obtains the more general relation ${\rm prob}_\pm(\varphi) = \frac{1}{2}(1 \pm {\rm Im}[\bra{\psi} \psi' \rangle \ee^{\ii \varphi}])$. A disorder average then yields $\overline{{\rm prob}}_\pm(\varphi) = \frac{1}{2}(1 \pm \frac{r+1}{2} \sin \varphi)$. We thus find that the purity loss of the disorder-averaged state quantifies the detrimental deviation of the beam splitter operation from the ideal case.

\section{Gap condition}

In the case of topological insulators, a prerequisite for backscattering-free propagation is that the wave packets remain, in momentum space, contained in the gap region, which is limited to a finite range in momentum space. An energy gap $\Delta$ between the two bulk bands then translates into a tolerable momentum range $\Delta/v$, which must not be exceeded due to disorder effects [Fig.~\ref{Fig:backscattering_free_transport}(b)].

To assess the disorder-induced momentum broadening, we evaluate the momentum variance $\langle (\Delta \hat{p})^2 \rangle \equiv \langle (\hat{p} -\langle \hat{p} \rangle)^2 \rangle$ for the solution~(\ref{Eq:master_equation_solution}), yielding
\begin{align} \label{Eq:momentum_variance}
\langle (\Delta \hat{p})^2 \rangle(t) = \langle (\Delta \hat{p})^2 \rangle_0 + \frac{4}{v^2} \int_{-\infty}^{\infty} \!\!\! \dd q \,\, G(q) \, \sin \! \left[ \frac{q v t}{2 \hbar} \right] ,
\end{align}
with $\langle (\Delta \hat{p})^2 \rangle_0$ the momentum variance of the initial state. In the case of Gaussian correlations and a Gaussian initial state, Eq.~(\ref{Eq:momentum_variance}) becomes $\langle (\Delta \hat{p})^2 \rangle(t) = \frac{\hbar^2}{4 \sigma^2} + \frac{2 C_0}{v^2} \left( 1 - \ee^{- (v t/\ell)^2} \right)$, which saturates after $t \gg \ell/v$. To remain at all times within the gap region, we thus impose the gap condition $\frac{\hbar^2}{4 \sigma^2} + \frac{2 C_0}{v^2} < \frac{\Delta^2}{v^2}$ (assuming that the wave packet is centered around the gap center). Note that the average momentum, which indicates the position of the wave packet in the band, is unaffected by the disorder; in case of the above Gaussian state, $\langle \hat{p} \rangle(t) = p_0$.

\section{Cyclic operation}

Edge modes usually propagate in a ring topology, where wave packets periodically return to their initial position and thus, on intermediate time scales, repeatedly encounter the same disorder realization. We can model this situation with a periodic correlation function, $C(x+L) = C(x)$, where $L$ denotes the ring circumference ($\ell \ll L$). This then results in a discrete momentum transfer distribution $G(q)$, such that, as expected, the purity and the momentum variance (indeed, the full momentum distribution, i.e., all moments) return to their initial, undisturbed values whenever $v t = n L$, with $n \in \mathbb{Z}$. This suggests to process states in the vicinity of their injection point.

\begin{figure}[htb]
	\begin{center}
	\includegraphics[width=0.99\columnwidth]{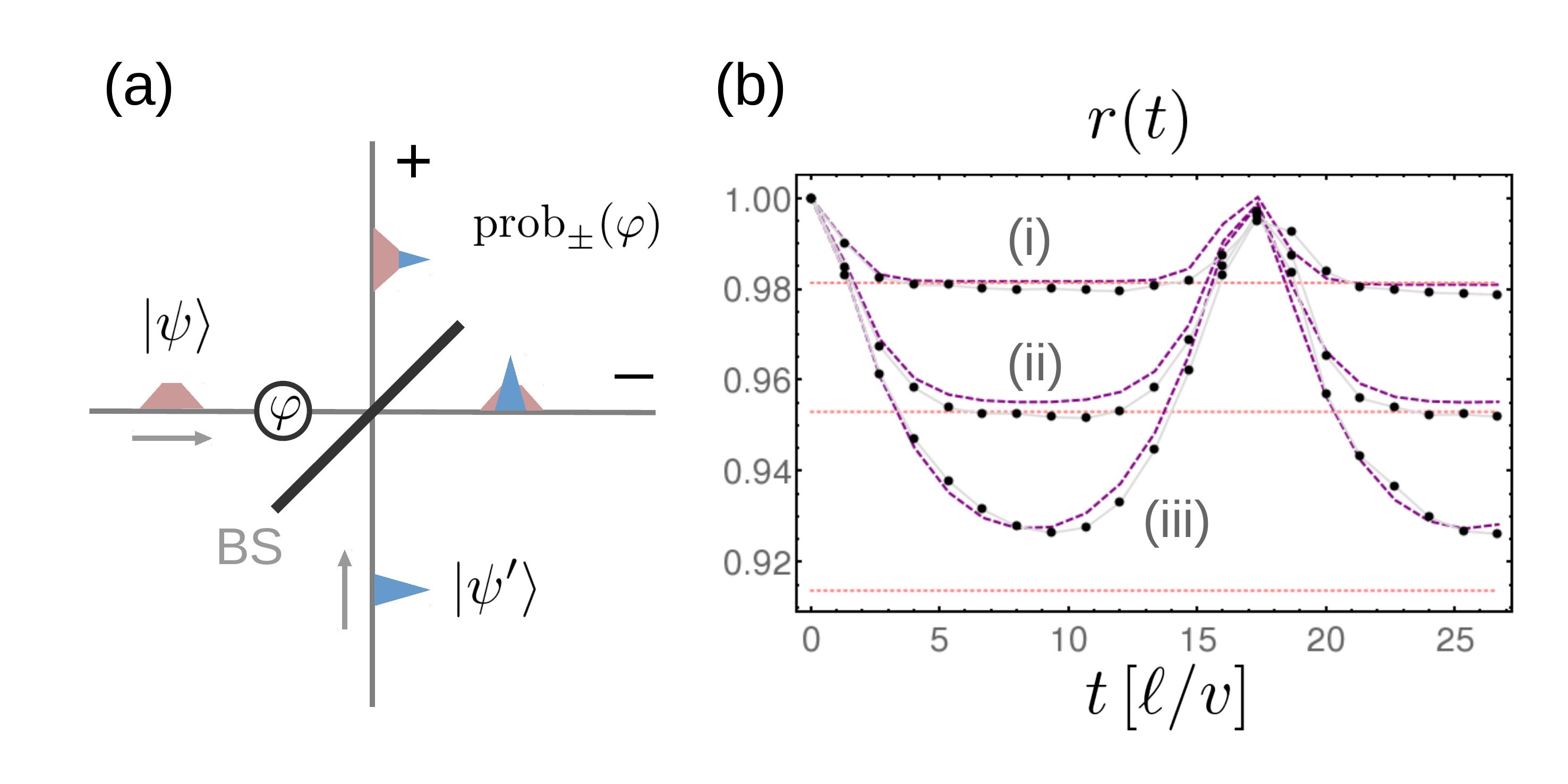}
	\caption{\label{Fig:purity_evolution} (a) Impact of disorder on the functioning of a beam splitter. Optimal operation, based on constructive and destructive interference in the output ports, presupposes identical input states $\ket{\psi}$ and $\ket{\psi'}$. If, due to different disorder histories of the state components, the wave packets differ, interference in the output ports is corrupted, resulting in detrimental leakage between the ports. (b) Purity evolution in a closed loop of circumference $L = 17 \ell$. Shown is the time evolution of the purity $r(t)$ for a Gaussian wave packet with (i) $\sigma = \ell$, (ii) $\sigma = 2 \ell$, and (iii) $\sigma = 10 \ell/3$. We compare the numerically exact evolution of the state, averaged over 500 disorder realizations (black dots), the prediction of the solution (\ref{Eq:master_equation_solution}) for periodic correlations (purple dashed curves), and the approximation (\ref{Eq:purity_evolution}) (red dotted lines). While in all cases the purity fully recovers when the state completes a full cycle, the intermediate purity loss scales with the wave packet extension. In case (iii), the purity decay is reversed before it reaches its saturation as predicted by Eq.~(\ref{Eq:purity_evolution}).}
	\end{center}
\end{figure}

In Fig.~\ref{Fig:purity_evolution}(b), we display the purity evolution for the case of locally Gaussian, periodic correlations, $C(x) = C_0 \sum_{n \in \mathbb{Z}} \ee^{-(x+n L)^2/\ell^2}$, with $C_0 = 7.5 \times 10^{-3} (\frac{v \hbar}{\ell})^2$ and $L=17 \ell$. To this end, we propagate Gaussian initial states with (i) $\sigma = \ell$, (ii) $\sigma = 2 \ell$, and (iii) $\sigma = 10 \ell/3$. We compare the numerically exact evolution, averaged over 500 disorder realizations, the prediction of the solution (\ref{Eq:master_equation_solution}), and the approximation (\ref{Eq:purity_evolution}). We find that our theory accurately predicts the evolution of the disorder-averaged state. In all three cases the purity undergoes full revivals after the state completes a full cycle. The intermediate purity loss, on the other hand, scales with the extension $\sigma$ of the wave packet. If the circumference $L$ becomes comparable to the wave packet extension, the revival sets in before the purity loss saturates, as shown in case iii).

\section{Validity discussion}

To discuss the validity of the perturbative master equation (\ref{Eq:master_equation_evaluated}) and its solution (\ref{Eq:master_equation_solution}), we exploit that the unitary time evolution of single realizations can be solved exactly in the considered scenario, $\rho_{\varepsilon}(x, x'; t) = \rho_0(x-v t, x'-v t) \, \ee^{-\frac{\ii}{\hbar} \int_{0}^{t} \dd t' (V_{\varepsilon}(x-v t) - V_{\varepsilon}(x'-v t))}$. The disorder-averaged state, $\overline{\rho}(x, x'; t) = \int \dd \varepsilon \, p_{\varepsilon} \rho_{\varepsilon}(x, x'; t)$, can then be written as
\begin{subequations}
\begin{align} \label{Eq:exact_solution}
\overline{\rho}(x, x'; t) =& \rho_0(x-v t, x'-v t) \\
&\times \left\{ 1 - \overline{F}_t(x-x') + \sum_{n=3}^{\infty} R_t^{(n)}(x, x') \right\} , \nonumber
\end{align}
with the disorder influence $\overline{F}_t(x)$ as in Eq.~(\ref{Eq:disorder_influence}) and the remainder terms
\begin{align} \label{Eq:rest_terms}
R_t&^{(n)}(x, x') = \\
&\frac{(-\ii/\hbar)^n}{n! \, v^n} \int \dd \varepsilon \, p_{\varepsilon} \left[ \int_{0}^{v t} \dd \tilde{x} \big( V_{\varepsilon}(x-\tilde{x}) - V_{\varepsilon}(x'-\tilde{x}) \big) \right]^n . \nonumber
\end{align}
\end{subequations}
Note that neglecting the remainder terms corresponds to a standard Born approximation of the state. Doing so, however, the solution would fail to be a manifestly well-defined quantum state. Instead, the solution~(\ref{Eq:master_equation_solution}) replaces the sum of remainder terms with $\sum_{n=2}^{\infty} \frac{1}{n!} [-\overline{F}_t(x-x')]^n$. Indeed, if the disorder potential is described by a multivariate normal distribution (with two-point correlations Gaussian or not), one obtains $R_t^{(2n+1)}(x, x') = 0$ and $R_t^{(2n)}(x, x') = \frac{1}{n!} [\overline{F}_t(x-x')]^n$, $n \in \mathbb{N}_+$. We thus find that, in that case, Eq.~(\ref{Eq:master_equation_solution}) is not only the exact solution of the perturbative master equation (\ref{Eq:master_equation_evaluated}), but, indeed, the exact solution of the disorder-averaged dynamics. In this sense, for general disorder distributions, Eqs.~(\ref{Eq:master_equation_evaluated}, \ref{Eq:master_equation_solution}) can be considered as approximations in the deviations from a multivariate normal distribution.

The spatial integral in the remanent terms~(\ref{Eq:rest_terms}) is, for $v t \gg \ell$, independent of the velocity $v$ and, due to the self-averaging of the disorder potential, takes (after disorder average) a plateau value when $|x-x'| \gg \ell$ and $v t \gg \ell$. Therefore, we can consider Eq.~(\ref{Eq:exact_solution}) an expansion in $1/v$, and the perturbative solution~(\ref{Eq:master_equation_solution}) becomes the more accurate the larger $v$. Note that, in contrast to propagation in a parabolic band, where the number of backscattering events limits the temporal validity of the perturbative master equation \cite{Gneiting2017quantum}, in the present case there is no such limitation.

\section{Conclusions}

We investigated the disorder effects on 1D systems which display backscattering-free, dispersionless propagation. This was achieved by establishing the Lindblad master equation (\ref{Eq:master_equation_evaluated}), which captures the time evolution due to the drift and its effects on the populations and the coherences of the disorder-averaged state, as revealed by its exact solution (\ref{Eq:master_equation_solution}). We found that the disorder-induced dephasing saturates, quantified by the purity evolution (\ref{Eq:purity_evolution}), which can be seen as a consequence of the conelike spatiotemporal decoherence behavior of the propagating wave packets. This bounded purity loss, which renders these systems attractive as carriers of quantum information, stands in stark contrast to the ongoing purity decay of the dispersively spreading wave packets in parabolic bands \cite{Gneiting2017quantum}. Moreover, we showed that, in loop configurations, the purity revives after full cycles.

The purity of the disorder-averaged state provides us with a statistically robust assessment of disorder effects on the functioning of devices. We exemplified this with our evaluation of disorder effects on beam splitters, which are among the essential processing units in the linear handling of spatial modes. Generally, we expect that our approach is useful in order to assess disorder effects on the functioning of quantum technologies that are based on backscattering-free, dispersionless propagation, including, e.g., photons \cite{Knill2001scheme, Aaronson2011computational} and graphene \cite{Ando1998impurity, Euen1999disorder, Allen2009honeycomb, Peres2010colloquium, Rozhkov2011electronic, Plotnik2014observation, Rozhkov2016electronic}. This also encompasses cases where the dispersion exhibits higher-power momentum contributions, if the linear component remains dominant.

The disorder-induced dephasing is accompanied by a broadening of the momentum distribution which, in the case of topological insulators, must not exceed the limits imposed by the bulk band gap on the backscattering-free propagation of the edge modes. This lead us to formulating a gap condition on the disorder-induced momentum broadening. While our approach is perturbative in the disorder potential, we showed that, in the case of Gaussian variates, it reproduces the exact ensemble dynamics, for arbitrary disorder strengths. Generally, its accuracy increases with increasing drift velocity.

\paragraph{Acknowledgments.}

This research was partially supported by the RIKEN iTHES Project, the MURI Center for Dynamic Magneto-Optics via the AFOSR Award No.~FA9550-14-1-0040, the Japan Society for the Promotion of Science (KAKENHI), the IMPACT program of JST, JSPS-RFBR Grant No.~17-52-50023, CREST Grant No.~JPMJCR1676, a Grant-in-Aid for Scientific Research (A), and a grant from the John Templeton Foundation.

\bibliography{literature}

\begin{thebibliography}{30}%
\makeatletter
\providecommand \@ifxundefined [1]{%
 \@ifx{#1\undefined}
}%
\providecommand \@ifnum [1]{%
 \ifnum #1\expandafter \@firstoftwo
 \else \expandafter \@secondoftwo
 \fi
}%
\providecommand \@ifx [1]{%
 \ifx #1\expandafter \@firstoftwo
 \else \expandafter \@secondoftwo
 \fi
}%
\providecommand \natexlab [1]{#1}%
\providecommand \enquote  [1]{``#1''}%
\providecommand \bibnamefont  [1]{#1}%
\providecommand \bibfnamefont [1]{#1}%
\providecommand \citenamefont [1]{#1}%
\providecommand \href@noop [0]{\@secondoftwo}%
\providecommand \href [0]{\begingroup \@sanitize@url \@href}%
\providecommand \@href[1]{\@@startlink{#1}\@@href}%
\providecommand \@@href[1]{\endgroup#1\@@endlink}%
\providecommand \@sanitize@url [0]{\catcode `\\12\catcode `\$12\catcode
  `\&12\catcode `\#12\catcode `\^12\catcode `\_12\catcode `\%12\relax}%
\providecommand \@@startlink[1]{}%
\providecommand \@@endlink[0]{}%
\providecommand \url  [0]{\begingroup\@sanitize@url \@url }%
\providecommand \@url [1]{\endgroup\@href {#1}{\urlprefix }}%
\providecommand \urlprefix  [0]{URL }%
\providecommand \Eprint [0]{\href }%
\providecommand \doibase [0]{http://dx.doi.org/}%
\providecommand \selectlanguage [0]{\@gobble}%
\providecommand \bibinfo  [0]{\@secondoftwo}%
\providecommand \bibfield  [0]{\@secondoftwo}%
\providecommand \translation [1]{[#1]}%
\providecommand \BibitemOpen [0]{}%
\providecommand \bibitemStop [0]{}%
\providecommand \bibitemNoStop [0]{.\EOS\space}%
\providecommand \EOS [0]{\spacefactor3000\relax}%
\providecommand \BibitemShut  [1]{\csname bibitem#1\endcsname}%
\let\auto@bib@innerbib\@empty
\bibitem [{\citenamefont {Hasan}\ and\ \citenamefont
  {Kane}(2010)}]{Hasan2010colloquium}%
  \BibitemOpen
  \bibfield  {author} {\bibinfo {author} {\bibfnamefont {M.~Z.}\ \bibnamefont
  {Hasan}}\ and\ \bibinfo {author} {\bibfnamefont {C.~L.}\ \bibnamefont
  {Kane}},\ }\bibfield  {title} {\enquote {\bibinfo {title} {Colloquium:
  {T}opological insulators},}\ }\href {\doibase 10.1103/RevModPhys.82.3045}
  {\bibfield  {journal} {\bibinfo  {journal} {Rev. Mod. Phys.}\ }\textbf
  {\bibinfo {volume} {82}},\ \bibinfo {pages} {3045--3067} (\bibinfo {year}
  {2010})}\BibitemShut {NoStop}%
\bibitem [{\citenamefont {Qi}\ and\ \citenamefont
  {Zhang}(2011)}]{Qi2011topological}%
  \BibitemOpen
  \bibfield  {author} {\bibinfo {author} {\bibfnamefont {X.-L.}\ \bibnamefont
  {Qi}}\ and\ \bibinfo {author} {\bibfnamefont {S.-C.}\ \bibnamefont {Zhang}},\
  }\bibfield  {title} {\enquote {\bibinfo {title} {Topological insulators and
  superconductors},}\ }\href {\doibase 10.1103/RevModPhys.83.1057} {\bibfield
  {journal} {\bibinfo  {journal} {Rev. Mod. Phys.}\ }\textbf {\bibinfo {volume}
  {83}},\ \bibinfo {pages} {1057--1110} (\bibinfo {year} {2011})}\BibitemShut
  {NoStop}%
\bibitem [{\citenamefont {Brendel}\ \emph {et~al.}(1999)\citenamefont
  {Brendel}, \citenamefont {Gisin}, \citenamefont {Tittel},\ and\ \citenamefont
  {Zbinden}}]{Brendel1999pulsed}%
  \BibitemOpen
  \bibfield  {author} {\bibinfo {author} {\bibfnamefont {J.}~\bibnamefont
  {Brendel}}, \bibinfo {author} {\bibfnamefont {N.}~\bibnamefont {Gisin}},
  \bibinfo {author} {\bibfnamefont {W.}~\bibnamefont {Tittel}}, \ and\ \bibinfo
  {author} {\bibfnamefont {H.}~\bibnamefont {Zbinden}},\ }\bibfield  {title}
  {\enquote {\bibinfo {title} {Pulsed energy-time entangled twin-photon source
  for quantum communication},}\ }\href {\doibase 10.1103/PhysRevLett.82.2594}
  {\bibfield  {journal} {\bibinfo  {journal} {Phys. Rev. Lett.}\ }\textbf
  {\bibinfo {volume} {82}},\ \bibinfo {pages} {2594--2597} (\bibinfo {year}
  {1999})}\BibitemShut {NoStop}%
\bibitem [{\citenamefont {Thew}\ \emph {et~al.}(2002)\citenamefont {Thew},
  \citenamefont {Tanzilli}, \citenamefont {Tittel}, \citenamefont {Zbinden},\
  and\ \citenamefont {Gisin}}]{Thew2002experimental}%
  \BibitemOpen
  \bibfield  {author} {\bibinfo {author} {\bibfnamefont {R.~T.}\ \bibnamefont
  {Thew}}, \bibinfo {author} {\bibfnamefont {S.}~\bibnamefont {Tanzilli}},
  \bibinfo {author} {\bibfnamefont {W.}~\bibnamefont {Tittel}}, \bibinfo
  {author} {\bibfnamefont {H.}~\bibnamefont {Zbinden}}, \ and\ \bibinfo
  {author} {\bibfnamefont {N.}~\bibnamefont {Gisin}},\ }\bibfield  {title}
  {\enquote {\bibinfo {title} {Experimental investigation of the robustness of
  partially entangled qubits over 11 km},}\ }\href {\doibase
  10.1103/PhysRevA.66.062304} {\bibfield  {journal} {\bibinfo  {journal} {Phys.
  Rev. A}\ }\textbf {\bibinfo {volume} {66}},\ \bibinfo {pages} {062304}
  (\bibinfo {year} {2002})}\BibitemShut {NoStop}%
\bibitem [{\citenamefont {Knill}\ \emph {et~al.}(2001)\citenamefont {Knill},
  \citenamefont {Laflamme},\ and\ \citenamefont {Milburn}}]{Knill2001scheme}%
  \BibitemOpen
  \bibfield  {author} {\bibinfo {author} {\bibfnamefont {E.}~\bibnamefont
  {Knill}}, \bibinfo {author} {\bibfnamefont {R.}~\bibnamefont {Laflamme}}, \
  and\ \bibinfo {author} {\bibfnamefont {G.~J.}\ \bibnamefont {Milburn}},\
  }\bibfield  {title} {\enquote {\bibinfo {title} {A scheme for efficient
  quantum computation with linear optics},}\ }\href@noop {} {\bibfield
  {journal} {\bibinfo  {journal} {Nature}\ }\textbf {\bibinfo {volume} {409}},\
  \bibinfo {pages} {46--52} (\bibinfo {year} {2001})}\BibitemShut {NoStop}%
\bibitem [{\citenamefont {Gneiting}\ \emph {et~al.}(2016)\citenamefont
  {Gneiting}, \citenamefont {Anger},\ and\ \citenamefont
  {Buchleitner}}]{Gneiting2016incoherent}%
  \BibitemOpen
  \bibfield  {author} {\bibinfo {author} {\bibfnamefont {C.}~\bibnamefont
  {Gneiting}}, \bibinfo {author} {\bibfnamefont {F.~R.}\ \bibnamefont {Anger}},
  \ and\ \bibinfo {author} {\bibfnamefont {A.}~\bibnamefont {Buchleitner}},\
  }\bibfield  {title} {\enquote {\bibinfo {title} {Incoherent ensemble dynamics
  in disordered systems},}\ }\href@noop {} {\bibfield  {journal} {\bibinfo
  {journal} {Phys. Rev. A}\ }\textbf {\bibinfo {volume} {93}},\ \bibinfo
  {pages} {032139} (\bibinfo {year} {2016})}\BibitemShut {NoStop}%
\bibitem [{\citenamefont {Kropf}\ \emph {et~al.}(2016)\citenamefont {Kropf},
  \citenamefont {Gneiting},\ and\ \citenamefont
  {Buchleitner}}]{Kropf2016effective}%
  \BibitemOpen
  \bibfield  {author} {\bibinfo {author} {\bibfnamefont {C.~M.}\ \bibnamefont
  {Kropf}}, \bibinfo {author} {\bibfnamefont {C.}~\bibnamefont {Gneiting}}, \
  and\ \bibinfo {author} {\bibfnamefont {A.}~\bibnamefont {Buchleitner}},\
  }\bibfield  {title} {\enquote {\bibinfo {title} {Effective dynamics of
  disordered quantum systems},}\ }\href@noop {} {\bibfield  {journal} {\bibinfo
   {journal} {Phys. Rev. X}\ }\textbf {\bibinfo {volume} {6}},\ \bibinfo
  {pages} {031023} (\bibinfo {year} {2016})}\BibitemShut {NoStop}%
\bibitem [{\citenamefont {Gneiting}\ and\ \citenamefont
  {Nori}(2017)}]{Gneiting2017quantum}%
  \BibitemOpen
  \bibfield  {author} {\bibinfo {author} {\bibfnamefont {C.}~\bibnamefont
  {Gneiting}}\ and\ \bibinfo {author} {\bibfnamefont {F.}~\bibnamefont
  {Nori}},\ }\bibfield  {title} {\enquote {\bibinfo {title} {Quantum evolution
  in disordered transport},}\ }\href {\doibase 10.1103/PhysRevA.96.022135}
  {\bibfield  {journal} {\bibinfo  {journal} {Phys. Rev. A}\ }\textbf {\bibinfo
  {volume} {96}},\ \bibinfo {pages} {022135} (\bibinfo {year}
  {2017})}\BibitemShut {NoStop}%
\bibitem [{\citenamefont {Chen}\ \emph {et~al.}()\citenamefont {Chen},
  \citenamefont {Gneiting}, \citenamefont {Lo}, \citenamefont {Chen},\ and\
  \citenamefont {Nori}}]{Chen2017simulating}%
  \BibitemOpen
  \bibfield  {author} {\bibinfo {author} {\bibfnamefont {H.-B.}\ \bibnamefont
  {Chen}}, \bibinfo {author} {\bibfnamefont {C.}~\bibnamefont {Gneiting}},
  \bibinfo {author} {\bibfnamefont {P.-Y.}\ \bibnamefont {Lo}}, \bibinfo
  {author} {\bibfnamefont {Y.-N.}\ \bibnamefont {Chen}}, \ and\ \bibinfo
  {author} {\bibfnamefont {F.}~\bibnamefont {Nori}},\ }\bibfield  {title}
  {\enquote {\bibinfo {title} {Simulating open quantum systems with
  {H}amiltonian ensembles and the nonclassicality of the dynamics},}\
  }\href@noop {} {\bibinfo  {journal} {arXiv:1703.09428}\ }\BibitemShut
  {NoStop}%
\bibitem [{\citenamefont {Lindblad}(1976)}]{Lindblad1976generators}%
  \BibitemOpen
\bibfield  {journal} {  }\bibfield  {author} {\bibinfo {author} {\bibfnamefont
  {G.}~\bibnamefont {Lindblad}},\ }\bibfield  {title} {\enquote {\bibinfo
  {title} {On the generators of quantum dynamical semigroups},}\ }\href@noop {}
  {\bibfield  {journal} {\bibinfo  {journal} {Commun. Math. Phys.}\ }\textbf
  {\bibinfo {volume} {48}},\ \bibinfo {pages} {119--130} (\bibinfo {year}
  {1976})}\BibitemShut {NoStop}%
\bibitem [{\citenamefont {Breuer}\ and\ \citenamefont
  {Petruccione}(2002)}]{Breuer2002theory}%
  \BibitemOpen
  \bibfield  {author} {\bibinfo {author} {\bibfnamefont {H.-P.}\ \bibnamefont
  {Breuer}}\ and\ \bibinfo {author} {\bibfnamefont {F.}~\bibnamefont
  {Petruccione}},\ }\href@noop {} {\emph {\bibinfo {title} {The Theory of Open
  Quantum Systems}}}\ (\bibinfo  {publisher} {Oxford Univ. Press, New York},\
  \bibinfo {year} {2002})\BibitemShut {NoStop}%
\bibitem [{\citenamefont {Jackiw}\ and\ \citenamefont
  {Rebbi}(1976)}]{Jackiw1976solitons}%
  \BibitemOpen
  \bibfield  {author} {\bibinfo {author} {\bibfnamefont {R.}~\bibnamefont
  {Jackiw}}\ and\ \bibinfo {author} {\bibfnamefont {C.}~\bibnamefont {Rebbi}},\
  }\bibfield  {title} {\enquote {\bibinfo {title} {Solitons with fermion number
  1/2},}\ }\href {\doibase 10.1103/PhysRevD.13.3398} {\bibfield  {journal}
  {\bibinfo  {journal} {Phys. Rev. D}\ }\textbf {\bibinfo {volume} {13}},\
  \bibinfo {pages} {3398--3409} (\bibinfo {year} {1976})}\BibitemShut {NoStop}%
\bibitem [{\citenamefont {Kossakowski}(1972)}]{Kossakowski1972quantum}%
  \BibitemOpen
  \bibfield  {author} {\bibinfo {author} {\bibfnamefont {A.}~\bibnamefont
  {Kossakowski}},\ }\bibfield  {title} {\enquote {\bibinfo {title} {On quantum
  statistical mechanics of non-{H}amiltonian systems},}\ }\href@noop {}
  {\bibfield  {journal} {\bibinfo  {journal} {Rep. Math. Phys.}\ }\textbf
  {\bibinfo {volume} {3}},\ \bibinfo {pages} {247--274} (\bibinfo {year}
  {1972})}\BibitemShut {NoStop}%
\bibitem [{\citenamefont {Manita}(1991)}]{Manita1991properties}%
  \BibitemOpen
  \bibfield  {author} {\bibinfo {author} {\bibfnamefont {A.~D.}\ \bibnamefont
  {Manita}},\ }\bibfield  {title} {\enquote {\bibinfo {title} {Properties of
  translation-invariant quantum-dynamical semigroups},}\ }\href@noop {}
  {\bibfield  {journal} {\bibinfo  {journal} {Theor. Math. Phys.}\ }\textbf
  {\bibinfo {volume} {89}},\ \bibinfo {pages} {1271--1281} (\bibinfo {year}
  {1991})}\BibitemShut {NoStop}%
\bibitem [{\citenamefont {Botvich}\ \emph {et~al.}(1991)\citenamefont
  {Botvich}, \citenamefont {Malyshev},\ and\ \citenamefont
  {Manita}}]{Botvich1991translation}%
  \BibitemOpen
  \bibfield  {author} {\bibinfo {author} {\bibfnamefont {D.~D.}\ \bibnamefont
  {Botvich}}, \bibinfo {author} {\bibfnamefont {V.~A.}\ \bibnamefont
  {Malyshev}}, \ and\ \bibinfo {author} {\bibfnamefont {A.~D.}\ \bibnamefont
  {Manita}},\ }\bibfield  {title} {\enquote {\bibinfo {title} {Translation
  invariant quantum master equation},}\ }\href@noop {} {\bibfield  {journal}
  {\bibinfo  {journal} {Helv. Phys. Acta}\ }\textbf {\bibinfo {volume} {64}},\
  \bibinfo {pages} {1072--1092} (\bibinfo {year} {1991})}\BibitemShut {NoStop}%
\bibitem [{\citenamefont {Holevo}(1995)}]{Holevo1995translation}%
  \BibitemOpen
  \bibfield  {author} {\bibinfo {author} {\bibfnamefont {A.~S.}\ \bibnamefont
  {Holevo}},\ }\bibfield  {title} {\enquote {\bibinfo {title} {On
  translation-covariant quantum {M}arkov equations},}\ }\href@noop {}
  {\bibfield  {journal} {\bibinfo  {journal} {Izv. Math.}\ }\textbf {\bibinfo
  {volume} {59}},\ \bibinfo {pages} {427} (\bibinfo {year} {1995})}\BibitemShut
  {NoStop}%
\bibitem [{\citenamefont {Gallis}\ and\ \citenamefont
  {Fleming}(1990)}]{Gallis1990environmental}%
  \BibitemOpen
  \bibfield  {author} {\bibinfo {author} {\bibfnamefont {M.~R.}\ \bibnamefont
  {Gallis}}\ and\ \bibinfo {author} {\bibfnamefont {G.~N.}\ \bibnamefont
  {Fleming}},\ }\bibfield  {title} {\enquote {\bibinfo {title} {Environmental
  and spontaneous localization},}\ }\href {\doibase 10.1103/PhysRevA.42.38}
  {\bibfield  {journal} {\bibinfo  {journal} {Phys. Rev. A}\ }\textbf {\bibinfo
  {volume} {42}},\ \bibinfo {pages} {38--48} (\bibinfo {year}
  {1990})}\BibitemShut {NoStop}%
\bibitem [{\citenamefont {Hornberger}\ and\ \citenamefont
  {Sipe}(2003)}]{Hornberger2003collisional}%
  \BibitemOpen
  \bibfield  {author} {\bibinfo {author} {\bibfnamefont {K.}~\bibnamefont
  {Hornberger}}\ and\ \bibinfo {author} {\bibfnamefont {J.~E.}\ \bibnamefont
  {Sipe}},\ }\bibfield  {title} {\enquote {\bibinfo {title} {Collisional
  decoherence reexamined},}\ }\href {\doibase 10.1103/PhysRevA.68.012105}
  {\bibfield  {journal} {\bibinfo  {journal} {Phys. Rev. A}\ }\textbf {\bibinfo
  {volume} {68}},\ \bibinfo {pages} {012105} (\bibinfo {year}
  {2003})}\BibitemShut {NoStop}%
\bibitem [{\citenamefont {Gneiting}\ \emph {et~al.}(2013)\citenamefont
  {Gneiting}, \citenamefont {Fischer},\ and\ \citenamefont
  {Hornberger}}]{Gneiting2013quantum}%
  \BibitemOpen
  \bibfield  {author} {\bibinfo {author} {\bibfnamefont {C.}~\bibnamefont
  {Gneiting}}, \bibinfo {author} {\bibfnamefont {T.}~\bibnamefont {Fischer}}, \
  and\ \bibinfo {author} {\bibfnamefont {K.}~\bibnamefont {Hornberger}},\
  }\bibfield  {title} {\enquote {\bibinfo {title} {Quantum phase-space
  representation for curved configuration spaces},}\ }\href@noop {} {\bibfield
  {journal} {\bibinfo  {journal} {Phys. Rev. A}\ }\textbf {\bibinfo {volume}
  {88}},\ \bibinfo {pages} {062117} (\bibinfo {year} {2013})}\BibitemShut
  {NoStop}%
\bibitem [{\citenamefont {Aaronson}\ and\ \citenamefont
  {Arkhipov}(2011)}]{Aaronson2011computational}%
  \BibitemOpen
  \bibfield  {author} {\bibinfo {author} {\bibfnamefont {S.}~\bibnamefont
  {Aaronson}}\ and\ \bibinfo {author} {\bibfnamefont {A.}~\bibnamefont
  {Arkhipov}},\ }\bibfield  {title} {\enquote {\bibinfo {title} {The
  computational complexity of linear optics},}\ }in\ \href@noop {} {\emph
  {\bibinfo {booktitle} {Proceedings of the {F}orty-{T}hird {A}nnual ACM
  Symposium on Theory of Computing}}}\ (\bibinfo {organization} {ACM, New
  York},\ \bibinfo {year} {2011})\ pp.\ \bibinfo {pages} {333--342}\BibitemShut
  {NoStop}%
\bibitem [{\citenamefont {Hong}\ \emph {et~al.}(1987)\citenamefont {Hong},
  \citenamefont {Ou},\ and\ \citenamefont {Mandel}}]{Hong1987measurement}%
  \BibitemOpen
  \bibfield  {author} {\bibinfo {author} {\bibfnamefont {C.~K.}\ \bibnamefont
  {Hong}}, \bibinfo {author} {\bibfnamefont {Z.~Y.}\ \bibnamefont {Ou}}, \ and\
  \bibinfo {author} {\bibfnamefont {L.}~\bibnamefont {Mandel}},\ }\bibfield
  {title} {\enquote {\bibinfo {title} {Measurement of subpicosecond time
  intervals between two photons by interference},}\ }\href {\doibase
  10.1103/PhysRevLett.59.2044} {\bibfield  {journal} {\bibinfo  {journal}
  {Phys. Rev. Lett.}\ }\textbf {\bibinfo {volume} {59}},\ \bibinfo {pages}
  {2044--2046} (\bibinfo {year} {1987})}\BibitemShut {NoStop}%
\bibitem [{\citenamefont {Tichy}\ \emph {et~al.}(2015)\citenamefont {Tichy},
  \citenamefont {Ra}, \citenamefont {Lim}, \citenamefont {Gneiting},
  \citenamefont {Kim},\ and\ \citenamefont {M{\o}lmer}}]{Tichy2015double}%
  \BibitemOpen
  \bibfield  {author} {\bibinfo {author} {\bibfnamefont {M.~C.}\ \bibnamefont
  {Tichy}}, \bibinfo {author} {\bibfnamefont {Y.-S.}\ \bibnamefont {Ra}},
  \bibinfo {author} {\bibfnamefont {H.-T.}\ \bibnamefont {Lim}}, \bibinfo
  {author} {\bibfnamefont {C.}~\bibnamefont {Gneiting}}, \bibinfo {author}
  {\bibfnamefont {Y.-H.}\ \bibnamefont {Kim}}, \ and\ \bibinfo {author}
  {\bibfnamefont {K.}~\bibnamefont {M{\o}lmer}},\ }\bibfield  {title} {\enquote
  {\bibinfo {title} {Double-fock superposition interferometry for differential
  diagnosis of decoherence},}\ }\href@noop {} {\bibfield  {journal} {\bibinfo
  {journal} {New J. Phys.}\ }\textbf {\bibinfo {volume} {17}},\ \bibinfo
  {pages} {023008} (\bibinfo {year} {2015})}\BibitemShut {NoStop}%
\bibitem [{\citenamefont {Ra}\ \emph {et~al.}(2017)\citenamefont {Ra},
  \citenamefont {Tichy}, \citenamefont {Lim}, \citenamefont {Gneiting},
  \citenamefont {M{\o}lmer}, \citenamefont {Buchleitner},\ and\ \citenamefont
  {Kim}}]{Ra2017reversed}%
  \BibitemOpen
  \bibfield  {author} {\bibinfo {author} {\bibfnamefont {Y.-S.}\ \bibnamefont
  {Ra}}, \bibinfo {author} {\bibfnamefont {M.~C.}\ \bibnamefont {Tichy}},
  \bibinfo {author} {\bibfnamefont {H.-T.}\ \bibnamefont {Lim}}, \bibinfo
  {author} {\bibfnamefont {C.}~\bibnamefont {Gneiting}}, \bibinfo {author}
  {\bibfnamefont {K.}~\bibnamefont {M{\o}lmer}}, \bibinfo {author}
  {\bibfnamefont {A.}~\bibnamefont {Buchleitner}}, \ and\ \bibinfo {author}
  {\bibfnamefont {Y.-H.}\ \bibnamefont {Kim}},\ }\bibfield  {title} {\enquote
  {\bibinfo {title} {Reversed interplay of quantum interference and which-way
  information in multiphoton entangled states},}\ }\href {\doibase
  10.1103/PhysRevA.96.023845} {\bibfield  {journal} {\bibinfo  {journal} {Phys.
  Rev. A}\ }\textbf {\bibinfo {volume} {96}},\ \bibinfo {pages} {023845}
  (\bibinfo {year} {2017})}\BibitemShut {NoStop}%
\bibitem [{\citenamefont {Ando}\ and\ \citenamefont
  {Nakanishi}(1998)}]{Ando1998impurity}%
  \BibitemOpen
  \bibfield  {author} {\bibinfo {author} {\bibfnamefont {T.}~\bibnamefont
  {Ando}}\ and\ \bibinfo {author} {\bibfnamefont {T.}~\bibnamefont
  {Nakanishi}},\ }\bibfield  {title} {\enquote {\bibinfo {title} {Impurity
  scattering in carbon nanotubes --absence of back-scattering--},}\ }\href
  {\doibase 10.1143/JPSJ.67.1704} {\bibfield  {journal} {\bibinfo  {journal}
  {J. Phys. Soc. Japan}\ }\textbf {\bibinfo {volume} {67}},\ \bibinfo {pages}
  {1704--1713} (\bibinfo {year} {1998})}\BibitemShut {NoStop}%
\bibitem [{\citenamefont {McEuen}\ \emph {et~al.}(1999)\citenamefont {McEuen},
  \citenamefont {Bockrath}, \citenamefont {Cobden}, \citenamefont {Yoon},\ and\
  \citenamefont {Louie}}]{Euen1999disorder}%
  \BibitemOpen
  \bibfield  {author} {\bibinfo {author} {\bibfnamefont {P.~L.}\ \bibnamefont
  {McEuen}}, \bibinfo {author} {\bibfnamefont {M.}~\bibnamefont {Bockrath}},
  \bibinfo {author} {\bibfnamefont {D.~H.}\ \bibnamefont {Cobden}}, \bibinfo
  {author} {\bibfnamefont {Y.-G.}\ \bibnamefont {Yoon}}, \ and\ \bibinfo
  {author} {\bibfnamefont {S.~G.}\ \bibnamefont {Louie}},\ }\bibfield  {title}
  {\enquote {\bibinfo {title} {Disorder, pseudospins, and backscattering in
  carbon nanotubes},}\ }\href {\doibase 10.1103/PhysRevLett.83.5098} {\bibfield
   {journal} {\bibinfo  {journal} {Phys. Rev. Lett.}\ }\textbf {\bibinfo
  {volume} {83}},\ \bibinfo {pages} {5098--5101} (\bibinfo {year}
  {1999})}\BibitemShut {NoStop}%
\bibitem [{\citenamefont {Allen}\ \emph {et~al.}(2010)\citenamefont {Allen},
  \citenamefont {Tung},\ and\ \citenamefont {Kaner}}]{Allen2009honeycomb}%
  \BibitemOpen
  \bibfield  {author} {\bibinfo {author} {\bibfnamefont {M.~J.}\ \bibnamefont
  {Allen}}, \bibinfo {author} {\bibfnamefont {V.~C.}\ \bibnamefont {Tung}}, \
  and\ \bibinfo {author} {\bibfnamefont {R.~B.}\ \bibnamefont {Kaner}},\
  }\bibfield  {title} {\enquote {\bibinfo {title} {Honeycomb carbon: a review
  of graphene},}\ }\href@noop {} {\bibfield  {journal} {\bibinfo  {journal}
  {Chem. Rev.}\ }\textbf {\bibinfo {volume} {110}},\ \bibinfo {pages}
  {132--145} (\bibinfo {year} {2010})}\BibitemShut {NoStop}%
\bibitem [{\citenamefont {Peres}(2010)}]{Peres2010colloquium}%
  \BibitemOpen
  \bibfield  {author} {\bibinfo {author} {\bibfnamefont {N.~M.~R.}\
  \bibnamefont {Peres}},\ }\bibfield  {title} {\enquote {\bibinfo {title}
  {Colloquium: The transport properties of graphene: An introduction},}\ }\href
  {\doibase 10.1103/RevModPhys.82.2673} {\bibfield  {journal} {\bibinfo
  {journal} {Rev. Mod. Phys.}\ }\textbf {\bibinfo {volume} {82}},\ \bibinfo
  {pages} {2673--2700} (\bibinfo {year} {2010})}\BibitemShut {NoStop}%
\bibitem [{\citenamefont {Rozhkov}\ \emph {et~al.}(2011)\citenamefont
  {Rozhkov}, \citenamefont {Giavaras}, \citenamefont {Bliokh}, \citenamefont
  {Freilikher},\ and\ \citenamefont {Nori}}]{Rozhkov2011electronic}%
  \BibitemOpen
  \bibfield  {author} {\bibinfo {author} {\bibfnamefont {A.~V.}\ \bibnamefont
  {Rozhkov}}, \bibinfo {author} {\bibfnamefont {G.}~\bibnamefont {Giavaras}},
  \bibinfo {author} {\bibfnamefont {Y.~P.}\ \bibnamefont {Bliokh}}, \bibinfo
  {author} {\bibfnamefont {V.}~\bibnamefont {Freilikher}}, \ and\ \bibinfo
  {author} {\bibfnamefont {F.}~\bibnamefont {Nori}},\ }\bibfield  {title}
  {\enquote {\bibinfo {title} {Electronic properties of mesoscopic graphene
  structures: Charge confinement and control of spin and charge transport},}\
  }\href@noop {} {\bibfield  {journal} {\bibinfo  {journal} {Phys. Rep.}\
  }\textbf {\bibinfo {volume} {503}},\ \bibinfo {pages} {77--114} (\bibinfo
  {year} {2011})}\BibitemShut {NoStop}%
\bibitem [{\citenamefont {Plotnik}\ \emph {et~al.}(2014)\citenamefont
  {Plotnik}, \citenamefont {Rechtsman}, \citenamefont {Song}, \citenamefont
  {Heinrich}, \citenamefont {Zeuner}, \citenamefont {Nolte}, \citenamefont
  {Lumer}, \citenamefont {Malkova}, \citenamefont {Xu}, \citenamefont
  {Szameit}, \citenamefont {Chen},\ and\ \citenamefont
  {Segev}}]{Plotnik2014observation}%
  \BibitemOpen
  \bibfield  {author} {\bibinfo {author} {\bibfnamefont {Y.}~\bibnamefont
  {Plotnik}}, \bibinfo {author} {\bibfnamefont {M.~C.}\ \bibnamefont
  {Rechtsman}}, \bibinfo {author} {\bibfnamefont {D.}~\bibnamefont {Song}},
  \bibinfo {author} {\bibfnamefont {M.}~\bibnamefont {Heinrich}}, \bibinfo
  {author} {\bibfnamefont {J.~M.}\ \bibnamefont {Zeuner}}, \bibinfo {author}
  {\bibfnamefont {S.}~\bibnamefont {Nolte}}, \bibinfo {author} {\bibfnamefont
  {Y.}~\bibnamefont {Lumer}}, \bibinfo {author} {\bibfnamefont
  {N.}~\bibnamefont {Malkova}}, \bibinfo {author} {\bibfnamefont
  {J.}~\bibnamefont {Xu}}, \bibinfo {author} {\bibfnamefont {A.}~\bibnamefont
  {Szameit}}, \bibinfo {author} {\bibfnamefont {Z.}~\bibnamefont {Chen}}, \
  and\ \bibinfo {author} {\bibfnamefont {M.}~\bibnamefont {Segev}},\ }\bibfield
   {title} {\enquote {\bibinfo {title} {Observation of unconventional edge
  states in `photonic graphene'},}\ }\href@noop {} {\bibfield  {journal}
  {\bibinfo  {journal} {Nat. Mater.}\ }\textbf {\bibinfo {volume} {13}},\
  \bibinfo {pages} {57--62} (\bibinfo {year} {2014})}\BibitemShut {NoStop}%
\bibitem [{\citenamefont {Rozhkov}\ \emph {et~al.}(2016)\citenamefont
  {Rozhkov}, \citenamefont {Sboychakov}, \citenamefont {Rakhmanov},\ and\
  \citenamefont {Nori}}]{Rozhkov2016electronic}%
  \BibitemOpen
  \bibfield  {author} {\bibinfo {author} {\bibfnamefont {A.V.}\ \bibnamefont
  {Rozhkov}}, \bibinfo {author} {\bibfnamefont {A.O.}\ \bibnamefont
  {Sboychakov}}, \bibinfo {author} {\bibfnamefont {A.L.}\ \bibnamefont
  {Rakhmanov}}, \ and\ \bibinfo {author} {\bibfnamefont {F.}~\bibnamefont
  {Nori}},\ }\bibfield  {title} {\enquote {\bibinfo {title} {Electronic
  properties of graphene-based bilayer systems},}\ }\href {\doibase
  https://doi.org/10.1016/j.physrep.2016.07.003} {\bibfield  {journal}
  {\bibinfo  {journal} {Phys. Rep.}\ }\textbf {\bibinfo {volume} {648}},\
  \bibinfo {pages} {1 -- 104} (\bibinfo {year} {2016})}\BibitemShut {NoStop}%
\end{thebibliography}%

\end{document}